\documentstyle[amssymb,preprint,aps]{revtex}
\begin{document}

\begin{center}
{\huge Reliable Teleportation of Ionic Motional States Through a Mapping
Process} \\[0.25cm]
{\bf \ N. G. de Almeida}$^1${\bf , C. J. Villas-B\^{o}as}$^1$, {\bf E. Solano%
}$^{2,3}${\bf , and } {\bf M. H. Y. Moussa}$^{1*}${\bf \ }\\[0pt]
$^1${\it Departamento de F\'{\i }sica, Universidade Federal de S\~{a}o
Carlos, Via Washington Luis, km 235, S\~{a}o Carlos 13565-905, SP, Brasil}

$^2${\it Instituto de F\'{\i }sica, Universidade Federal do Rio de
Janeiro,Caixa Postal 68528, 21945-970 Rio de Janeiro, RJ, Brazil\\[0pt]
}$^3${\it Secci\'{o}n F\'{\i }sica, Departamento de Ciencias, Pontificia
Universidad Cat\'{o}lica del Per\'{u}, Apartado 1761, Lima, Peru}\\[0pt]
\end{center}

%\vspace{0.5cm} 
%% BEGIN ABSTRACT %%%%%%%%%%%%%%%%%%%%%%%%%%%%%%%%%%%%%%%%%%%%%%%%%%

\begin{center}
{\bf Abstract}
\end{center}

We show how to teleport reliably an arbitrary superposition of $n=0$ and $n=1
$ vibrational number state between two distant ions. This is done by first
mapping the vibrational state to be teleported into the internal degrees of
freedom of a given ion. Then we handle with the internal superposition state
following Bennett's original protocol and a recently proposed technique for
teleportation of ionic internal states [quant-ph/9903029]. Finally, the
teleportation of the vibrational state is achieved by reversing the mapping
process in the receiver ion. We remark that as in the teleportation of
cavity field and atomic states, the teleportation of vibrational states is
100\% successful for an ideal process.

\noindent PACS number(s): \ 03.65.Bz, 03.67.-a, 42.50.Dv

\noindent $^{*}${\it E-mail: miled@power.ufscar.br}

Recently, the teleportation phenomena proposed by Bennett et al. \cite
{bennett} has been demonstrated experimentally for photon polarized states 
\cite{dik,boschi} (discrete variables) and for the wave function of a single
mode of the electromagnetic field (continuous variables)\cite{kimble}. There
is a number of proposals for teleporting an arbitrary atomic state \cite
{davi,norton} using high-${\em Q}$ cavities and for teleporting cavity field
states \cite{moussa,scully}. In a recent work Solano et al. proposed a
reliable teleportation of internal states of single trapped ions \cite
{solano1,solano2}. In this work we present a reliable teleportation of an
arbitrary superposition of $n=0$ and $n=1$ vibrational number state between
two distant ions. To this end, we have developed an external$%
\leftrightharpoons $internal state mapping process. First we map the
external state to be teleported into the internal degrees of freedom of a
given ion (ion $1$) by means of a specific unitary operation. Then, we use
Ref.\cite{solano2} in order to proceed with Bennett's original protocol for
teleportation handling with the ionic internal states. After the required
Alice's joint measurement on ions $1$ and $2$ (which together with ion $3$
composes the quantum channel) and the communication of the measurement
result to Bob, by means of a classical channel, he finally reverses the
mapping process on ion $3$ recovering from its internal degrees of freedom
the teleported vibrational state.

The arbitrary vibrational state of ion $1$, $\alpha \left| 0\right\rangle
_1+\beta \left| 1\right\rangle _1$, is easily generated, in trap $A$ (Alice
station), by a sequence of a carrier and an anti-JC pulses. Such state of
ion $1$ is then mapped into its internal state by another single anti-JC
pulse producing an unitary evolution $U(t)$. Assuming that ion $1$ is
prepared in its excited electronic state $\left| e\right\rangle _1$, the
operator $U(t)$, after a $\pi $ pulse, leads to the mapping 
\begin{equation}
\left| e_1\right\rangle \left[ \alpha \left| 0_1\right\rangle +\beta \left|
1_1\right\rangle \right] \stackrel{U(t)}{\rightarrow }\left[ \alpha \left|
e_1\right\rangle -i%
%TCIMACRO{\limfunc{e}}
%BeginExpansion
\mathop{\rm e}%
%EndExpansion
\nolimits^{i\theta }\beta \left| g_1\right\rangle \right] \left|
0_1\right\rangle ,  \label{1}
\end{equation}
where $\theta $ is a known phase associated to the laser field. Due to this
unitary evolution, Alice does not need to know, as in Bennett's original
protocol, the vibrational state to be teleported despite that such state has
to be prepared in Alice's trap $A$. 

Now we consider, in trap $B$ (Bob station), the preparation of the maximally
entangled internal state of ions $2$ and $3$ using the technique developed
in Ref. \cite{solano1}. In this work the four Bell states are generated in a
deterministic way without modifying the vibrational state by means of the
Hamiltonian 
\begin{equation}
H=\hbar \left| \Omega \right| \left[ \left( S_{+j}S_{+k}%
%TCIMACRO{\limfunc{e}}
%BeginExpansion
\mathop{\rm e}%
%EndExpansion
\nolimits^{2i\phi }-S_{+j}S_{-k}%
%TCIMACRO{\limfunc{e}}
%BeginExpansion
\mathop{\rm e}%
%EndExpansion
\nolimits^{i\phi _0}-1/2\right) 
%TCIMACRO{\limfunc{e}}
%BeginExpansion
\mathop{\rm e}%
%EndExpansion
\nolimits^{i\varphi }+h.c.\right] \text{,}  \label{2}
\end{equation}
where $S_{+m}=\left| \uparrow _m\right\rangle \left\langle \downarrow
_m\right| $ and $S_{-m}=S_{+m}^{\dagger }$ are the electronic raising and
lowering operators acting on ion $m$, $\phi $ is the effective phase of each
Raman laser pair, taking as equal, and $\phi _o$ is the known phase
difference associated with the equilibrium separation of the two ions as
defined in Ref. \cite{solano1}. The effective coupling constant does not
depend on the vibrational quantum state and reads $\left| \Omega \right| =$ $%
2\Omega _o\eta ^2/\delta $, where $\Omega _o$ stands for each of the two
Raman effective Rabi frequencies. $\eta $ accounts for the Lamb-Dicke
parameter and $\delta $ for the detuning of the pair of Raman lasers
interacting dispersively with the ion pair. Let us consider that ions $2$
and $3$ were previously cooled to the Lamb-Dicke regime and prepared in
their electronic ground-state. After interacting with both Raman laser pairs
during a time interval $\tau =\pi /(4\left| \Omega \right| )$, following the
dynamics generated by hamiltonian (\ref{1}), the ions get prepared in the
EPR electronic state 
\begin{equation}
\left| \Phi _{23}\right\rangle =\frac 1{\sqrt{2}}\left( \left|
g_2,g_3\right\rangle -i%
%TCIMACRO{\limfunc{e}}
%BeginExpansion
\mathop{\rm e}%
%EndExpansion
\nolimits^{2i\phi _B}\left| e_2,e_3\right\rangle \right) \text{,}  \label{3}
\end{equation}
where $\phi _B$ is the effective phase of both Raman lasers pairs in trap $B$%
. Following the steps outlined in Ref. \cite{solano1}, one finally get the
electronic state of ion $1$, described by Eq.(\ref{1}), teleported to the
internal state of ion $3$. Basically, we have to move adiabatically the ion $%
2$ from trap $B$ to trap $A$, approaching it to ion $1$. By applying the
pair of Raman laser during the time $\tau =\pi /(4\left| \Omega \right| )$,
following the evolution due to hamiltonian (\ref{1}), the product state $%
\left[ \alpha \left| e_1\right\rangle -i%
%TCIMACRO{\limfunc{e}}
%BeginExpansion
\mathop{\rm e}%
%EndExpansion
\nolimits^{i\theta }\beta \left| g_1\right\rangle \right] \otimes \left|
\Phi _{23}\right\rangle $ evolves as 
\begin{eqnarray}
&&\frac 12\left\{ -i%
%TCIMACRO{\limfunc{e}}
%BeginExpansion
\mathop{\rm e}%
%EndExpansion
\nolimits^{2i\phi _A}\left| e_1,e_2\right\rangle \left[ \alpha \left|
g_3\right\rangle -i%
%TCIMACRO{\limfunc{e}}
%BeginExpansion
\mathop{\rm e}%
%EndExpansion
\nolimits^{2i\left[ \phi _B-\phi _A+\theta /2\right] }\beta \left|
e_3\right\rangle \right] \right.   \nonumber \\
&&+\left| g_1,g_2\right\rangle \left[ \alpha \left| g_3\right\rangle +i%
%TCIMACRO{\limfunc{e}}
%BeginExpansion
\mathop{\rm e}%
%EndExpansion
\nolimits^{2i\left[ \phi _B-\phi _A+\theta /2\right] }\beta \left|
e_3\right\rangle \right]   \nonumber \\
&&+\left| e_1,g_2\right\rangle \left[ -i%
%TCIMACRO{\limfunc{e}}
%BeginExpansion
\mathop{\rm e}%
%EndExpansion
\nolimits^{i\theta }\beta \left| g_3\right\rangle +%
%TCIMACRO{\limfunc{e}}
%BeginExpansion
\mathop{\rm e}%
%EndExpansion
\nolimits^{i\left[ 2\phi _B+\phi _0\right] }\alpha \left| e_3\right\rangle
\right]   \nonumber \\
&&\left. -i%
%TCIMACRO{\limfunc{e}}
%BeginExpansion
\mathop{\rm e}%
%EndExpansion
\nolimits^{-i\phi _0}\left| g_1,e_2\right\rangle \left[ i%
%TCIMACRO{\limfunc{e}}
%BeginExpansion
\mathop{\rm e}%
%EndExpansion
\nolimits^{i\theta }\beta \left| g_3\right\rangle +%
%TCIMACRO{\limfunc{e}}
%BeginExpansion
\mathop{\rm e}%
%EndExpansion
\nolimits^{i\left[ 2\phi _B+\phi _0\right] }\alpha \left| e_3\right\rangle
\right] \right\} .  \label{4}
\end{eqnarray}
Now, Alice only needs to make a measurement on the disentangled basis for
ions $1$ and $2$ through a well known fluorescence technique in order to
project the internal state of ion $3$ in one of the four possibilities shown
in Eq. [4]. The result of the measurement is sent to Bob by means of a
classical channel. For the accomplishment of the teleportation of the
vibrational state in (\ref{1}), Bob just needs to reverse the mapping
process described by the evolution $U(t)$ in the receiver ion $3$ in trap $B$%
. For doing this it is necessary that after the transport of ion $2$ to trap 
$A$ the vibrational state of ion $3$ must be taken to its fundamental level.
Current efforts are being considered for cooling the ionic vibrational state
without modifying the internal degrees of freedom by means of an auxiliary
ion inside an accumulator \cite{wineland}. Depending on the result of
Alice's measurement, $\left| e_1,e_2\right\rangle $, $\left|
g_1,g_2\right\rangle $ or $\left| e_1,g_2\right\rangle $, $\left|
g_1,e_2\right\rangle $, Bob has to apply a $\pi $ JC (${\cal U}(t)$) or a $%
\pi $ anti-JC ($U(t)$) pulse, respectively, in order to get the ion $3$
finally in one of the states 
\begin{mathletters}
\label{5}
\begin{eqnarray}
&&\left[ \alpha \left| g_3\right\rangle -i%
%TCIMACRO{\limfunc{e}}
%BeginExpansion
\mathop{\rm e}%
%EndExpansion
\nolimits^{2i\left[ \phi _B-\phi _A+\theta /2\right] }\beta \left|
e_3\right\rangle \right] \left| 0_3\right\rangle \stackrel{{\cal U}(t)}{%
\rightarrow }\left| g_3\right\rangle \left[ \alpha \left| 0_3\right\rangle -%
%TCIMACRO{\limfunc{e}}
%BeginExpansion
\mathop{\rm e}%
%EndExpansion
\nolimits^{2i\left[ \phi _B-\phi _A+\left( \theta +\chi \right) /2\right]
}\beta \left| 1_3\right\rangle \right] ,  \label{5a} \\
&&\left[ \alpha \left| g_3\right\rangle +i%
%TCIMACRO{\limfunc{e}}
%BeginExpansion
\mathop{\rm e}%
%EndExpansion
\nolimits^{2i\left[ \phi _B-\phi _A+\theta /2\right] }\beta \left|
e_3\right\rangle \right] \left| 0_3\right\rangle \stackrel{{\cal U}(t)}{%
\rightarrow }\left| g_3\right\rangle \left[ \alpha \left| 0_3\right\rangle +%
%TCIMACRO{\limfunc{e}}
%BeginExpansion
\mathop{\rm e}%
%EndExpansion
\nolimits^{2i\left[ \phi _B-\phi _A+\left( \theta +\chi \right) /2\right]
}\beta \left| 1_3\right\rangle \right] ,  \label{5b} \\
&&\left[ -i%
%TCIMACRO{\limfunc{e}}
%BeginExpansion
\mathop{\rm e}%
%EndExpansion
\nolimits^{i\theta }\beta \left| g_3\right\rangle +%
%TCIMACRO{\limfunc{e}}
%BeginExpansion
\mathop{\rm e}%
%EndExpansion
\nolimits^{i\left[ 2\phi _B+\phi _0\right] }\alpha \left| e_3\right\rangle
\right] \left| 0_3\right\rangle \stackrel{U(t)}{\rightarrow }%
%TCIMACRO{\limfunc{e}}
%BeginExpansion
\mathop{\rm e}%
%EndExpansion
\nolimits^{i\left[ 2\phi _B+\phi _0\right] }\left| e_3\right\rangle \left[
\alpha \left| 0_3\right\rangle -i%
%TCIMACRO{\limfunc{e}}
%BeginExpansion
\mathop{\rm e}%
%EndExpansion
\nolimits^{i\left[ -2\phi _B-\phi _0+\theta +\widetilde{\theta }\right]
}\beta \left| 1_3\right\rangle \right] ,  \label{5c} \\
&&\left[ i%
%TCIMACRO{\limfunc{e}}
%BeginExpansion
\mathop{\rm e}%
%EndExpansion
\nolimits^{i\theta }\beta \left| g_3\right\rangle +%
%TCIMACRO{\limfunc{e}}
%BeginExpansion
\mathop{\rm e}%
%EndExpansion
\nolimits^{i\left[ 2\phi _B+\phi _0\right] }\alpha \left| e_3\right\rangle
\right] \left| 0_3\right\rangle \stackrel{U(t)}{\rightarrow }%
%TCIMACRO{\limfunc{e}}
%BeginExpansion
\mathop{\rm e}%
%EndExpansion
\nolimits^{i\left[ 2\phi _B+\phi _0\right] }\left| e_3\right\rangle \left[
\alpha \left| 0_3\right\rangle +i%
%TCIMACRO{\limfunc{e}}
%BeginExpansion
\mathop{\rm e}%
%EndExpansion
\nolimits^{i\left[ -2\phi _B-\phi _0+\theta +\widetilde{\theta }\right]
}\beta \left| 1_3\right\rangle \right] ,  \label{5d}
\end{eqnarray}
where $\chi $ and $\widetilde{\theta }$ are known phases associated to the
lasers pulses leading to the reverse mapping evolutions ${\cal U}(t)$ and $%
U(t)$. Let us consider that the effective phase of both Raman laser pairs in
traps $A$ and $B$ were previously set so that $\phi _A=\phi _B=\pi -\phi _0/2
$. By having Alice communicated to Bob the phase $\theta $ together with her
measurement result, all Bob has to do is to adjust his laser phases $\chi $
or $\widetilde{\theta }$ in order to get the original vibrational state of
ion $1$ teleported to ion $3$.

We note that Bob could has proceeded to the teleportation of the electronic
state in Eq. (\ref{1}), as done in Ref. \cite{solano2}. In this case, after
knowing from Alice the result of her measurements on ions $1$ and $2$, Bob
has first to apply a $\pi $ rotation (around the {\it x}, {\it y}, or {\it z}
axis, if necessary), on the state of ion $3$, in order to get exactly the
teleported electronic state. After that, Bob proceeds to the reverse mapping
process which now requires only the $\pi $ anti-JC pulse, but at the
expenses of increasing the time interval required to achieve teleportation.

Let us now account for some sensitive points in the present experimental
scheme. Despite that the deterministic generation of both arbitrary one ion
internal and motional state has already been reported \cite{meekhof}, the
generation of the required EPR pair in Eq.(\ref{3}) is still an experimental
challenge. However, the EPR state as well as the required pairs of Raman
laser beams may be achieved with current technology as detailed in \cite
{solano2}. The transportation of ions from one to another trap without
affecting their internal state is another sensitive point which has been a
matter of present investigation by the group at NIST \cite{private}.
Finally, we mention that the vibrational state of a trapped ion can be
examined through the evolution of its internal states \cite{meekhof}, which
can be readout with nearly 100\% detection efficiency as reported in Ref. [%
\cite{nagourney}].

The process of mapping and reverse-mapping, together with the manipulation
of the internal degrees of freedom, guarantees a theoretical 100\%
probability of success for teleporting an arbitrary vibrational state. It is
worth to stress that the typical lifetime of a vibrational state is of the
order of milliseconds, which makes the mapping process attractive for
teleporting short-living states. In fact, being the typical lifetime of
internal ionic state of the order of minutes, we have time enough to
accomplish the teleportation of the motional state through the reverse
mapping process.

\begin{center}
{\bf Acknowledgments}
\end{center}

We wish to thank the support from CNPq and FAPESP, Brazil.

\end{mathletters}


\begin{references}
\bibitem{bennett}  C. H. Bennett, G. Brassard, C. Cr\'{e}peau, R. Jozsa, A.
Peres, and W. Wootters, Phys. Rev. Lett. {\bf 70}, 1895 (1993).

\bibitem{dik}  D. Bouwmeester, J.-W. Pan, K. Mattle, M. Eibl, H. Weinfurter,
and A. Zeilinger, Nature {\bf 390}, 575 (1997).

\bibitem{boschi}  D. Boschi, S. Branca, F. De Martini, L. Hardy, and S.
Popescu, Phys. Rev. Lett. {\bf 80}, 1121 (1998).

\bibitem{kimble}  A. Furusawa, J. L. Sorensen, S. L. Braunstein, C. A.
Fucks, H. J. Kimble, and E. S. Polzik, Science {\bf 282, }706 (1998).

\bibitem{davi}  L. Davidovich, N. Zagury, M. Brune, J. M. Raimond, and S.
Haroche, Phys. Rev. A {\bf 50}, R895 (1994).

\bibitem{norton}  N. G. Almeida, L. P. Maia, C. J. Villas-B\^{o}as, and M.
H. Y. Moussa, Phys. Lett. A {\bf 241}, 213 (1998).

\bibitem{moussa,scully}  M. H. Y. Moussa, Phys. Rev. A {\bf 54}, 4661
(1996); M. S. Zubairy, Phys. Rev. A {\bf 58}, 4368 (1999). 

\bibitem{solano1}  E. Solano, R. L. Matos Filho, and N. Zagury, Phys. Rev. A 
{\bf 59}, R2539 (1999).

\bibitem{solano2}  E. Solano, C. L. Cesar, R. L. de Matos Filho, and N.
Zagury, submitted for publication (available in quant-ph/9903029).

\bibitem{wineland}  D. J. Wineland, C. R. Monroe, W. M. Itano, D. Leibfried,
B. E. King, and D. M. Meekhof, NIST J. Research {\bf 103} (3), 259 (1998)
(Avaiable at http://nvl.nist.gov/pub/nistpubs/jres/jres.html).

\bibitem{meekhof}  D. M. Meekhof, C. Monroe, B. E. King, W. M. Itano, and D.
J. Wineland, Phys. Rev. Lett {\bf 76}, 1796 (1006).

\bibitem{private}  D. J. Wineland, private communication.

\bibitem{nagourney}  W. Nagourney, J. Sandberg, and H. Dehmelt, Phys. Rev.
Lett {\bf 56}, 2797 (1986).
\end{references}
\end{document}